# Critical Comparative Analysis and Recommendation in MAC Protocols for Wireless Mesh Networks using Multi-Objective Optimization and Statistical Testing

Ankita Singh[1], Sudhakar Singh[1*], Shiv Prakash[1]

## Abstract

Wireless Mesh Network (WMN) is surely one of the prominent networks in the modern era which is widely used in numerous evolving applications, viz. broadband home networking (BHN), community and neighbourhood networks (CNN), coordinated network management (CNM), and intelligent transportation systems (ITS), etc. It is a wireless network (WN) with multi-hop formed by many fixed wireless mesh routers (WMR) that are connected wirelessly with a mesh-alike backbone arrangement. In the IEEE 802.11s network, the node selection, scalability, stability, density of the nodes, mobility of the nodes, transmission power, and routing are major issues that WMN suffers. In this paper, a critical review of MAC protocols and their Quality of Service (QoS) parameters for WMN is presented to attain a better understanding of MAC protocols. Furthermore, the critical comparative analysis and recommendation of MAC procedures for WMN using Multi-objective optimization and statistical testing framework are performed. This framework is used for the analysis and recommendation of different protocols available for QoS parameters.

**Keywords:** Wireless mesh network (WMN), MAC protocols, Non-overlapping interference, channel assignment, Multi-objective optimization, statistical testing.

*Corresponding author

Ankita Singh
ankita.singh87@gmail.com

Sudhakar Singh
sudhakar@allduniv.ac.in

Shiv Prakash
shivprakash@allduniv.ac.in

[1] Department of Electronics and Communication, University of Allahabad, Prayagraj, India

# 1. Introduction

A Wireless Mesh Network (WMN) [1] is a special kind of multi-hop wireless network (WN) designed by many fixed wireless-mesh routers that are connected wirelessly via a mesh-like backbone arrangement. It is substituting the Wireless Infrastructure Networks (WIN) in several areas due to higher flexibility and cost-effectiveness [2]. Moreover, it affords network access for both mesh and general users over mesh routers and mesh users. Bridge functions are formed and the mesh router also delivers the marginal mobility and makes the backbone of WMNs. WMN suffers from various issues [3], such as scalability, stability, interoperability, node selection, and load balancing in a simple routing protocol for the IEEE 802.11s network [4]. The use of the node relation distance contributes to a problem of path instability. Since the IEEE 802.11 has been established, many problems have been overcome but still, certain issues and challenges are remained open. The overall issues in WMN's traffic strategy and the research challenges in the IEEE 802.11 network are discussed in this paper. The WMN can also be used for TVTA (Threat and Vulnerability Testing Assessment), Ad-Hoc Networks, and additionally for sensor networks. Nevertheless, the protocols available in these networks are not directly working in WMN's Therefore due to this research gap, new techniques and protocols specific to the WMN need to be developed. There is a lot of literature present in the areas of routing, scalability, and security in the IEEE 802.11s network.

The literature presently addresses enhanced performance in Ad-Hoc networks, but the nodes in Ad-Hoc Wireless Networks [5] compared with WMNs are unlikely to use Protocols of WMNs. The wireless Ad-Hoc networks are commonly used for the exchange of data between the users' group with a single-hop or multiple-hops, while WMN users are aiming to communicate in wired/wireless networks with public internet servers. WMNs are therefore supposed to support their services to server from client through multiple-hops. WMNs vary considerably from networks of Ad-Hoc and thus no protocols for Ad-Hoc network can be utilized to operate over many hops, as is the case for WMNs. The traffic flows are directed to public internet servers (root nodes) originating from WMN customers (leaf nodes), while Ad-Hoc networking traffic flows are peer to peer.

Medium Access Control (MAC) is a key component that guarantees the successful operation of WMN. Wireless Ad-Hoc networks [6] want users to switch very often, while users of WMN can move slowly. Protocols of Ad-Hoc network are therefore altered to take account of limited user versatility but access multiple-hop connectivity. It is also noted that MRs (Mesh Routers) are not mobile while MCs (Mesh Clients) are often expected to move. Since the MRs can be operated by AC while the MCs are operated by the battery. Therefore, a different fixed cap for MCs and MRs may be a smart idea; therefore Ad-Hoc network protocols are not used specifically for WMNs. IEEE 802.11 DCF (Distributed coordination function) the currently being used MAC protocol in WLANs, is observed by Hosseinabadi and Vaidya [7] as a consequence of high system idle times or collisions when it comes to channel use, channel access time, and the system throughput.

The materials used in literature review are collected by using the search keywords as given in Table 1.

Table **1.** Keywords used in collecting literature

| **Keyword Scenario 1** | **Keyword Scenario 2** |
|---|---|
| MAC protocols | Wireless Mess Networks |
| MAC protocol | Wireless Mess Network |
| MAC + protocols | Wireless+ Mess+ Network |
| MAC + protocol | Wireless+ Mess+ Networks |
| 'MAC protocols' | 'Wireless Mess Networks' |
| "MAC protocols" | 'Wireless Mess Network' |
| 'MAC protocol' | "Wireless Mess Networks" |
| "MAC protocol" | "Wireless Mess Network" |
| 'MAC + protocols' | 'Wireless+ Mess+ Network' |
| 'MAC + protocol' | 'Wireless+ Mess+ Networks' |
| "MAC + protocols" | "Wireless+ Mess+ Network" |

The organization of the paper after the introduction is as follows. Section 2 describes the MAC issues and MAC protocols. A detailed literature survey with critical analysis is discussed in section 3. Section 4 discusses the metric, multi-objective optimization, and statistical framework, and finally, in Section 5, the paper is concluded with future work.

## 2. Wireless Mesh Network

A WMN is another form of WN that can be connected within a hop distance by each node. The Knots (meter sites) over a hop distance are connected via the appropriate routing protocol. Various WMN architectures may be available: backbone which is either server client-based or hybrid. There are mesh routers in the architecture of Backbone. Figure 1 shows the general architecture of a WMN.

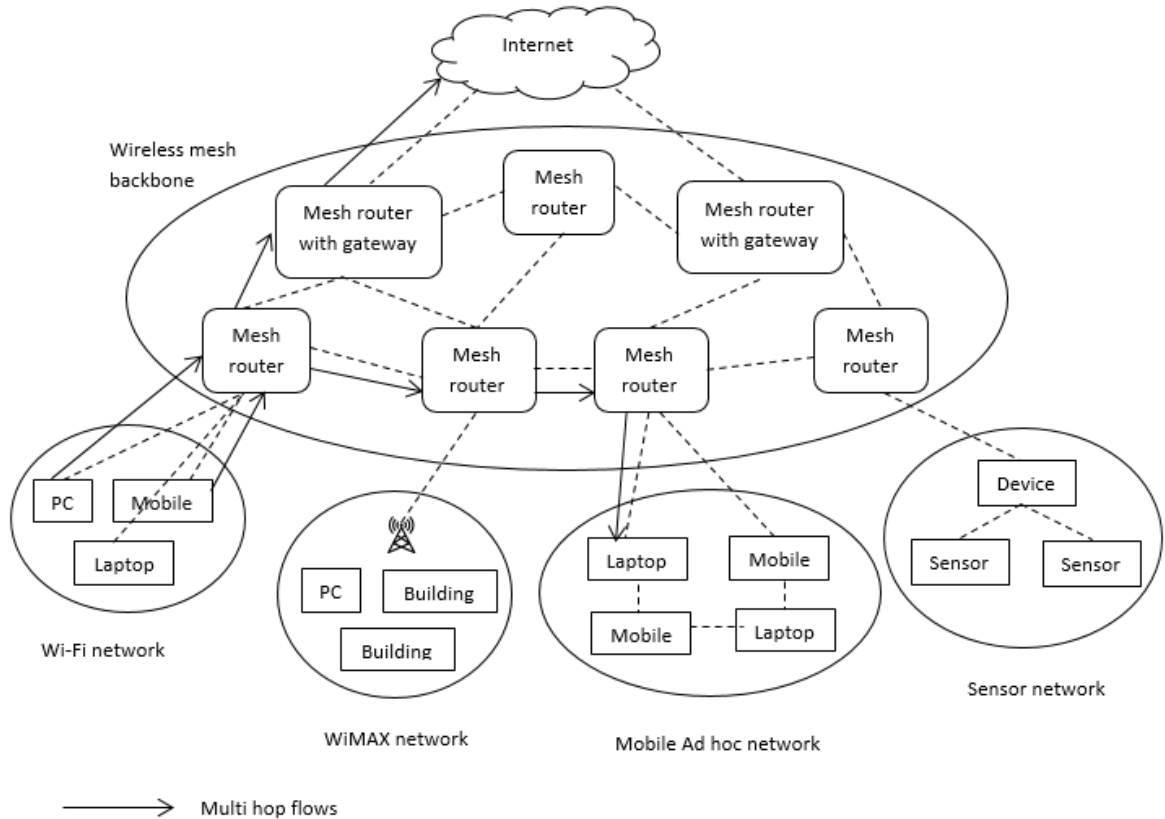

Figure 1: Architecture of a wireless mesh network

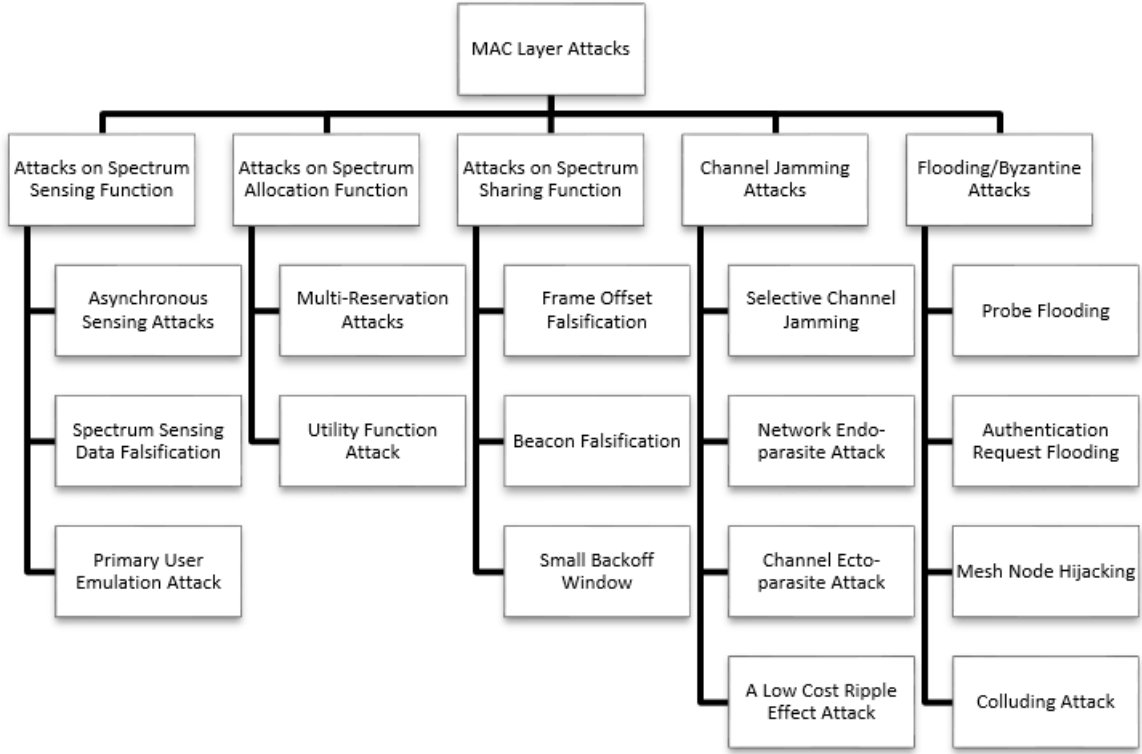

Figure 2: Taxonomy of MAC layer attacks

Figure 2 shows the classification of MAC layer security attacks. One important challenging issue in WMN is the capacity of operative throughput which can be given to the clients. But due to the broadcasting nature of the WMN, signals transmitted from dissimilar devices over a similar channel causes collision so, there is data loss. Therefore, multiple access methods viz. a time division multiple access (TDMA), a frequency multiple access (FDMA) are needed to coordinate the transmissions over the different channels. To minimize the interference, the transmission by devices may be through the dissimilar non-overlapping channels which are provisioned. The high network capacity of a multi-radio WMN can be achieved by how several channels are allocated efficiently to each radio to form a WMN with minimum interference which is termed as the channel assignment problem. This problem is NP-hard [8].

## 3. Comprehensive Review of Literature

In the conventional wireless network, the unknown terminal issue occurs, since two nodes are outside their carriers during CSMA (Carrier-sense multiple access) and they all want to connect with a similar node. To solve this conventional issue, the CTS (clear to send)/RTS (request to send) handshaking technique is implemented in order to prevent a collision [9], before the transmission of data. The omni-directional antenna typically distributes the electromagnetic force of a wireless signal over a vast space field, but the intended receiver absorbs only a very small part, thus theoretically restricting the total power and effectiveness. Furthermore, the omni-directional antenna has some common issues, such as multi-path deterioration, delay propagation, and interference from the co-channel. At present, it makes beam-forming antennas accessible to cable communications systems [10] by providing low-cost computing ability and developing new algorithms to process signals from simple antenna arrays. Intelligent antennas, composed of multi-straight antennas (MBA), are widely used in beam shaping antennas. By controlling temporal speed among antenna arrays with the DSP (digital signal processing) units.

Bao et al. have presented a channel access scheduling protocol (CASP) known as distributed receiver-oriented multiple access (ROMA) for antenna-based ad-hoc network. It specifies the number of contacts for activation in any slot utilizing only 2-hop topologies. Unlike spontaneous signal scanning or handshakes for communication targets on-demand access schemes. Its key advantage is having multiple beams that form directional antennas in both reception and transmission, which increases network efficiency and delay. With the aim to obtain channel access schedules for the Node, ROMA holds a Neighbour aware Contention Resolution (NCR). In each context, contention to the common resource is settled by the priorities allocated to the entities on basis of the context number and the identifiers, and the maximum preferences are given to get into the mutual resource without disagreement [4]. Every ROMA connection weights the data flow requirement for the connection on the topology of the ROMA network and the weight is automatically evaluated using the link's head to track traffic requirements or to receive requests for bandwidth from applications in the upper layer. The nodes and connections are targets dependent on their identities and the new Timeline ROMA is a recipient-oriented multi-access protocol that uses the ability to build multi-beam adaptive array antennas with different beams. Due to the up-to-date details about node and bandwidth distribution in the two-hop district, ROMA specifies if a node is a WMN transmitter or receiver, and which links for receipt or propagation can be activated during a particular time $t$. The ROMA protocol must settle on the working input links of every node in acceptance mode prior to the actual link activation on the transmitters.

Lal et al. [11] have suggested a new spatial division MAC multiple access protocol for the ad-Hoc WNs. Later, these ad-hoc wireless networks with directional antennas cannot change dramatically because they do not allow the nodes to carry out multiple simultaneous receptions/transmissions. It has been suggested that MAC layer protocol which makes use of multiple space division access to receive multiple packages from spatially separated transmitter nodes utilizing the directional reception property. The multi-access space division is split into two Time Division Duplex (TDD) structures among transmission and receipt. It uses parallelism and increases the throughput at a node in the receiving process, which allows the future transmitters to be synced to a recipient. Because any communication is determined by a possible recipient, nodes transmitting their transmissions to another cannot be synchronized. The system follows the recipient-initiated method to achieve reception time synchronization.

Smart antennas may use properly designed network protocols for the high level, like MAC protocols to exploit their superior capability. However, relative to standard MAC protocols, there are already many interface problems. Originally the conventional network logs are configured for the operation of nodes with omni-directional antennas that cannot communicate with the smart multi-beam antennas underpinned by omni-directional antennas and that can deteriorate the overall efficiency well below that obtained without careful control [12]. So, as to take advantage of the possible benefits in the WMNs it is therefore important to explore revolutionary protocols, particularly in the MAC layer. Lal et al. [13], have presented an assessment of the medium access control (MAC) efficiency in wireless LAN for multi-beam antenna nodes. They have used a broad-azimuth beam intelligent antenna system composed of a variety of beam antenna systems, and both CSMA and Slotted Aloha's one-hop performance was measured and simulated. The paper also discussed the synchronization problem for many CSMA beams. These findings indicate that CSMA, considering the lack of output due to beam syncing, is a good alternative in heavy-offered load conditions for nodes that are multi-beam smart antennas. In comparison, CSMA offers a consistent unity of approaches to transmission throughput and invariantly adjusts the load provided. However, slotted aloha efficiency decreases significantly below the optimal loads available, while slotted Aloha can produce higher peaks with available load's narrow range if more shifted beams are used.

The CSMA analytical model is simple CSMA, not handshake, not ACK. The sense of the carrier is handled by an interface on a separate channel, consisting of a series of narrow belt tones than that used for transmitting data. The model considers that the approximate recipient number of beams is mapped in an omnidirectional tone and that the confidential tone should be fixed to hit all the members of a recipient beam, thus being as low as possible to minimize the risk of error. With some changes, the analytical slotted aloha model is identical to the one of CSMA. The latest smart multi-beam antennas may be widely divided into three types: versatile array, switched multi-beam and MIMO (multiple-input multiple-output) links antennas [14]. Sundaresan et al. [15] have introduced the ad hoc network media access control with multiple inputs multiple outputs (MIMO). Through concurrently distributing many independent data streams over the single path, MIMO links can provide incredibly high spectrally efficient multi-way streaming. An entirely new MAC protocol is required for the specific features of MIMO links combined with many primary optimization considerations. The authors demonstrate a centralized algorithm called SCMA (sparse code multiple access), which integrates the core concerns of optimizing the architecture of a stream-controlled media access.

The main advantage of space multiplexing is to maximize the system's ability. These modes are based on the discovery of the receptor's overheating problem using the unified stream regulated medium access protocol to ad hoc network connection. In the case of pure stream control, there is a certain sub-set of links within the network which contributes to the loss of recipient overload. In order to track the bottleneck connections overflows, they use the timetable as controlled by non-stream. The connections can effectively be separated from all planning factors and only with separate contention areas, the scheduling algorithm can be used with pure Stream Control. With the growing prevalence of local wireless connectivity, there is a huge demand that data sharing among access points and terminals or base stations increase throughput and energy consumption. Wireless networks are historically configured for single-hop transmission to base stations or to access points via WLAN. But, the increase in wireless network installations in practice inspired the concept of WMN [16], expanding wireless reach, enhancing maximum capacity, and allowing the self-configuration of the network. A network of networking nodes arranged into a topology of mesh, which is also a type of a WLAN ad-hoc network, is the WMN.

A complementary beam-forming was introduced by Choi et al. [17] who suggested two new techniques to handle the problem of complementary field. A region in which certain network stations are unable to detect directional signals (beams) that often cause concealed beam problems: "Subspace Complementary Beam-forming (SCBF)" and "Complementary Superposition Beam-forming (CSBF)". The SCBF utilizes dummy data to ensure the regulated level of the energy obtained by downlink channels in both directions. In the meantime, CSBF requires it to send data that is also used in the supplementary beam. Use this approach to get "broadcast" information to passive nodes within the network by utilizing active nodes to share user-specific data. Multiple accesses are not implemented in space-division, the additional beam forming effects can be done easily by optimizing only one antenna unit transmission power. The key concept behind CBF (cubic feet) is that unintended users require far less power to reliably find the existence of transmission that is sufficient generally for the appropriate decoding of transmitted packet. The SCBF technique produces a smooth beam pattern in the directions of the 'otherwise secret beam' such that power is guaranteed to be greater than a certain amount in both directions and positions. In CSBF, the SCBF vectors are paired with a linear composition. The side lobe is expanded, without crossing with the desired beams, by adjusting one of the SCBF's column vectors. The ultimate goal of this architecture is to maximize the likelihood of the channel operation being heard by all the nodes so that the transmission is interrupted. This ensures that the power obtained can surpass the threshold at every place in the service area.

Each antenna technology has its advantages and drawbacks. The switched multi-beam antennas, which have been deployed in many true applications, are very basic and commercially available [18]. In the wireless network, the secret terminal issue is much more severe. In order to synchronize shared data between the adjacent nodes, particularly with regard to mobility, it is important to provide a much more complicated MAC protocol. Mobility may not be a significant issue for conventional AP (access point) with omnidirectional antennas. In the WMN, APs with multi-beam antennas have a non-negligible problem where every node is connected directly to a beam field. When a mobile node transfers into another from the medium access control, one beam sector, in particular the download media access control, is strongly influenced. It takes an appropriate location update algorithm to keep location information fresh and relatively costly, thus making the WMN more complicated [18]. One of the goals of a multi-beam-forming antenna, which was first defined in accordance with simple MAC directives [19], [20], [21], [22], is to use spatial reusability. Deafness can happen, which is so far the biggest concern in a wireless network. However, as a transmitter attempts to connect with a source, because the source is moved to a direction other than the transmitter, this attempted transmission process fails. The expected recipient is not capable of transmitting the signal of the transmitter, unlike all omnidirectional antennas. Therefore, the transmitter appears deaf.

Multi-beam smart antenna WMN has recently gained a lot of popularity because it performs better and more effectively. But MAC protocols like IEEE 802.11, which are still common for confrontation, are less efficient for multi-beam antennas. Wang et al. [18] in the year 2017 suggested a way to enhance the usefulness of a multi-beam access points medium access control for WLANs. In order to resolve these difficult issues carefully and improve collaboration efficacy, the authors introduced a new MAC protocol. Their architecture also takes into account the reverse compatibility, with a multi-beam access point being opened transparently by an IEEE 802.11 terminal. In the paper, the authors agree that their device is equipped with multiple beam access points, a single frequency channel and handheld all-direction nodes, and a distributed solution of the IEEE 802.11 DCF layer for MAC layers. The MAC Protocol data link is split into two sorts: Medium Access Control Downlink and Medium Access Uplink. A timing structure to allow several handshakes (sequentially or overlapping) prior to parallel-collision-free data transmission would be the fundamental concept of their MAC protocol. Both uplink nodes in the queues contain uplink packets during the confinement time for uplink medium access control. They build a conflict settlement mechanism to make it possible to win several nodes so that the winning nodes are collision-free. The "collision-free" is thus simply a form where the nodes that are winning won't collide when the winning nodes transmit or get ACK data from the AP concurrently. So, as to be collision-free, multiple winning nodes can also be included in various beam sectors. The access point must compete with the channel like an ordinary node for WMNs. In this situation, the AP will send RTR (receiver initiated MAC) messages with greater priority to exchange RTS (sender initiated MAC) messages via mobile terminals. The information of the location of every node should be buffered by the WMN connection point to monitor the downlink medium access. They provide both static and mobile network situations. The position data collected during the affiliation process are statically buffered in the static scenario. The details on the beam location of each individual node, while present in the mobile scenario, are kept cached by AP and can be modified reactively when data needs to be shared or proactively modified using periodic polling/testing triggered by the AP.

On-demand media connectivity in multiple beam smart antennas supported multi-hop Wi-Fi networks was suggested by Jain et al. [23]. This special function is not feasible with the conventional MAC protocols for the directional, omnidirectional, and one ray antennas built on the DCF mechanism, or with the simultaneous transmitting or receiving of the multiple ray antennas. This article proposes to backward IEEE 802.11 DCF-compatible new MAC hybrid (HMAC) protocol. This protocol allows simultaneous packet receiving (CPR) and simultaneous packet transmission (CPT) on multiple antennas.

In the HMAC protocol [23], the architecture takes into account a large azimuth-switched smart beam antenna composed of a multiple beam antenna array (MBAA) that can measure the accurate arrival angle of the input signal. The HMAC, a cross-layer protocol, utilizes both physical layer information and network to run it. The latest features of HMAC include the channel entry system, deafness- and contention-reduction algorithm, jump back, and priority roles systems for throughput enhancement. Every node of the HMAC retains neighbour information within the hybrid network allocation vector (HNAV) table and a beam table (BT) that stores back-off time and two variables of Boolean for every beam. In the case of the HMAC, each node preserves neighbour information. HMAC uses a hybrid solution depending on the algorithm to minimize deafness (AMD) to eradicate the mitigated surface problem. HMAC uses an operating sending estimate algorithm (Rivest, Shamir, and Adleman (RSA) algorithm) to overcome the channel containment issue. The node retains a probability of transmission for each neighbouring node in the HNAV table, and the current approximation of the transmission number in that beam is reciprocal to this probability. Based on the packets in its buffer, HMAC can also execute priority position shifts between both the receiver and the transmitter roles. The rule is that, when the data packet is on the queue, the node takes precedence over the transmitting mode.

Shihab et al. [24] have presented a wireless ad hoc network asynchronous directional MAC protocol. The existing MAC protocols presume that both directional and omnidirectional nodes will work but, using all modes, the asymmetry-in-gain problem will arise at the same time. The authors also suggested a MAC protocol, where the receivers and senders run in a one-way fashion, and the saturation of the ad hoc networks is also supported by DtD (directional-to-directional) MAC. The protocol of DtD MAC is completely distributed and does not need coordination, avoids the issue of asymmetry gain, and decreases the consequences of irritation or crashes. DtD MAC protocol is dependent on the idea that nodes transmit caching the position information about their adjacent nodes that are later utilized to decide in which direction RTS (DRTS) messages should be sent in the first place. Whereas the idle nodes, as potential receptors suggest, scan omnidirectional antennas constantly through their antenna parts, potential recipients lock into the corresponding path and respond through a directional CTS (DCTS) message until they hear a DRTS programmed for them. The angle of arrival (AoA) of every message it eavesdrops to decide the position of the destination node is determined and cached by each node. In order to decide the recipient's likely path, the node sent checks its AoA cache using a machine learning algorithm before sensing the media. The nodes which overhear ongoing communication accordingly shall refrain from interrupting continuous communication in this way by defining their Directional Network Allocation Vector (DNAV).

Wireless networks are mostly made up of mesh subscribers, gateways, and mesh routers [25]. Owing to the higher antenna advantages, longer propagation range, improved spatial reuse, and less disturbance among multi-beam antennas, the utilization of multi-beam (intelligent antennas) in WMNs have gained greater publicity [18]. The use of multi-straight smart antennas, particularly for WMNs in wireless LAN, therefore is of great interest. WMN, and its uses in the war and disaster relief climate, have consequently been increasingly interesting [26]. In [27], addressed end-to-end allotment issues for cognitive WMN radio in order to create a rational and effective trade-off. Two reasonable bandwidth assignment issues have been established, which are (a) the Lexicographical Max–Min (LMM) fair Bandwidth Allocation (LMMBA) problem and (b) the Max-Min equal maximum bandwidth allocation (MMBA) problem. Linear Programming (LP) for both problems is also recommended for optimal and quick heuristic (MMBA & LMMBA) algorithms. Numerical studies have shown that the performance of the LMMBA and MMBA algorithms is very similar to the acceptable maximum performance. LMMBA algorithm can also be used to achieve the fairest division of bandwidth.

The new concept [28] that put forward in response to the issue of multicast interference in multi-radio multi-channel wireless mesh networks (MR-MC WMNs) between multicast trees was suggested. The network can use the dynamic model traffic, i.e., multi-cast requests arrive dynamically without past visibility into potential attempts. Every network node functions as a Transit Access Point (TAP) and is tuned to non-overlap channels for one or more radios. It is also shown that in MR-MC WMNs, it was NP-hard to locate the Multicast Tree with reduced transmission costs, which is the Lowest Cost Multicast Tree (MCMT). A model known as Wireless Closest Terminal Branching (WCTB) has the same issue integrated with an ILP (integer-linear programming) approach, which includes a polynomial algorithm similar to time-for an optimum solution. The implementation of an effective relation planning algorithm is of utmost importance in order to have high throughput for a WMN as [1] observed. Three channel scheduling algorithms were suggested, based on the spatial TDMA (time division multiple access) networks, namely (a) the Network Connection Chart Model, (b) the Network Interference and Noise Ratio Signal, and (c) the Network Chart Model and the receiver threshold conditions.

We observed that interference and colliding substantially reduce mesh network performance using MAC protocols based on containment like IEEE 802.11 DCF. If transmissions are planned, considerably higher throughput is attainable. The typical methods of estimating optimum schedules are however intractable computationally. The device works poorly if the interference with the co-channel is ignored and if it happens. The paper [29] suggested a functional methodology for estimating optimization problem-setting optimum schedules that finds the same solution as the entire issue for a low-dimensional optimization issue. It is observed that there is often such a small concern. The final algorithm focuses easily arithmetically or geometrically, based on the goal of optimizing the minimum performance or optimizing proportional fair performance, where there is a minimum of all network flows. It has also been determined that the presented algorithms can determine optimal schedules for 2048-node networks within several minutes and 128-node networks within a few seconds [29]. It has been noted that the Multi-hop WMNs often suffer interference-related connection faults, dynamic barriers, and bandwidth requirement applications. This leads to a significant deterioration in efficiency in WMNs or involves costly manual network management to restore in real-time. In the article [30], an ARS (Autonomous network Reconfiguration System) is added, enabling a multi-radio WMN to extract autonomously network output defects from local connections. ARS produces required modifications on station assignments and local radio to recover from faults by the use of station and radio diversity in WMNs. Next, the device co-configures network configurations between local network routers based on the configuration changes thus produced [30].

The utility-based services distribution in backbone WMNs is discussed by Liu et al. [31]. A WMN will accept multi-hop communications of many contending links, as opposed to single-hop cell networks. It ensures that the distribution of resources is deliberately configured to produce the best performance. In the proportionally equal planning for WMNs, the clique-based approach of successful spatial reuse is implemented. A CBPFS (clique-based proportionally fair scheduling), a new algorithm has been presented by them. The plan was made to improve cellular network service to multi-speed wireless users [32]. Wi-Fi was designed to provide the wired network with last mile broadband internet connectivity. The last hop in the wired backhaul is WLAN, which is needed to link the wired and wireless network (WLAN) Access Points (AP). The APs only support single-hop wireless communication. The implementation of wired backhaul is required to provide wireless network connectivity to unserved areas. IEEE 802.11s fulfils the wireless communication gap for many hops, minimising the expense of creating and configuring the wired rear wireless link. The 802.11s has eliminated WLAN infrastructure dependence and allows for arbitrary topology to form autonomous WMNs. Also, for other networks the WMN 802.11s can provide a transparent backhaul. It also introduces a new MAC protocol that would allow scheduled access to the wireless medium based on distributed reserve.

A multi-hop MAC Protocol method based on 802.11s for WMNs was submitted by Carrano et al. [33]. The prosed mesh network is integrated into the connection layer and rather than IP addresses, it is based on its mechanisms on MAC addresses. This role helps the creation and production of new layer 2 multi-hop CPU-free network systems. The current layer 3 transmission is based on the WMN implementations, since these networks are based on layer 3 transmission based on 802.11 (WLAN), 802.3 (Ethernet), and 802.16 (WPAN). The running time of the mesh router would be reduced considerably if the connection packets for the wireless mesh router could be transmitted by the data layer. Due to layer 2 propagation, thereby greatly minimize transmission delays across the multi-hop networks. In response to challenges such as the unidirectional connection problem, heterogeneous exposed terminal problem, heterogeneous hidden terminal problem, and performance enhancement of the network, Huang et al. [34] suggested a cross-layer solution for mesh access networks. It has been suggested a thorough debate on the main source of CSMA-based network instability, here a network is considered to be stable only if (almost surely) every node's tail stays finite. Two variables that influence stability have been identified: the scale of the network and the stolen effect. In light of greedy sources, Foster's theorem has proven to be robust in three-hop networks only if they allow for the robbery effect. Furthermore, four-hop networks were found to be unreliable on the contrary (even without the issue of the robbery). In order to resolve the instability problem, it was suggested to use a distributed flow control system known as the EZ-Flow that is completely compliant with the IEEE 802.11 standard. In the article, it is also shown that when the hops among users are equal to or smaller than 3 hops but become unreliable when they are more than THREE hops, this WMN operates typically without degradation of service [35].

Researchers from academia and industry were drawn to the advantages of a WMN multi-beam antenna, which contributes to quick marketing as well as various standardization efforts [36]. The smart antenna system will relay data from or to the other nodes simultaneously, allowing directional antenna nodes or many omni-antennas to greatly increase the performance. This kind of hidden terminal issue is produced by the lack of canal state data during beam development. When a node is active, it appears deaf to other directions such that valuable information of control can be missed amid that period. The secret terminal issue is triggered when a nearby node is not receiving the channel reservation packets shared by a transmission receiver pair, like CTS and RTS. In this situation, the receiver does not aware about the imminent contact between the specific transmitter-receiver pair, such that a transmission that triggers the collision will later be triggered [36]. The Access Point (AP) and mobile station's finite range is a standard local wireless network. The AP is typically much more efficient and physically less limited than a Full Function Unit (FFU) mobile station. The AP typically supplies numerous intelligent antennas to improve network performance by space reuse [37].

Chou et al. [37], have presented MAC protocols, called M-HCCA, which reap the advantages of multi-beam AP antennas. This protocol improves not only the general ability of wireless Internet connectivity; it also promotes QoS (quality of service) and energy usage for each mobile station. A mobile station can detect the presence of APs through active scanning or passive scanning to access Wi-Fi services. The AP usually functions during the Contentious Free Time (CFP) in multi-tray antenna mode (except in an area dense with multi-paths). M-HCCA uses the recognition algorithm for tree-splitting during the collision resolution process for the resolution of collision issues. The stack is to be used by this algorithm to perform the dimension splitting tree's pre-order traversal.

Babich et al. [38] IEEE 802.11, the asynchronous network with the architectural conditions of adaptive antenna arrays, allow several peer-to-peer communications simultaneously and suggest two new access schemes for multi-packet communication (MPC). The designed MPC threshold access (TAMPC) is dependent on an effective loading threshold by a single node; the other system is based upon and on the adoption of the parity control codes of the low density, the accurate calculations of the signal-to-interference ratio (SIR). Both systems use the information collected by any node during network behaviour surveillance that is acceptable for scenarios that are heterogeneous and clustered. Its setting takes account of the various antenna systems of legacy and non-legacy nodes' coexistence. Both devices are also structured to be return compliant with the standard of IEEE 802.11.

Kuperman et al. [39] proposed an uncoordinated, non-coordinated ALOHA-like MAC Multi-beam Directional System random-access strategy that asymptotically achieves the network functionality. Each communication node acts separately in its setting. MB-URAM (Multi-beam Uncoordinated Random Access MAC) is not used to synchronize the time slots or transmissions with any of the booking messages like scheduling or RTS / CTS. Also, a scheduled MAC solution is able to reach asymptotically optimal performance for the proposed protocol. The authors considered realistic aspects of MB-URAM efficiency including bandwidth, beam-width power limits, and the packet error rate during the simulation.

Wang et al. [40] suggested a multi-beam directional antenna MAC protocol where the own control channel has been provided for every beam-sector and communication between various beam-sectors is autonomous. It uses the DNAV to record the institution's operations. It uses the DNAV vector. Later on, the global assignment technique is utilized by the protocol to delegate directional contact lines, after detecting all the multi-trunk antenna sectors. The entire protocol has several steps: activation of the network, contention of the channel, and communication of data. In each beam field, these steps are sequential.

Hong et al. [41] have explored 5G wireless networking multi-beam antenna technologies. In order to maintain an increased signal-to-interference plus range, a high transmission rate, increasing energy, spectral quality, and scalable beam. Multi-beam antennas are very exciting to be the crucial infrastructure for beam-forming and large 5G-enhancing MIMOs. The paper addressed in detail the application of multi-tray antennas on 5G environments. Medjo et al. [42] submitted the maximum capacity for multi-beam antennas' MPP/Multi-packet Receivability can be unlocked such the end-to-end latency of ad hoc networks can be significantly decreased. The authors established a structured delay reduction optimization model and observed that an optimum delay is achieved by creating linkages in such a way as to optimize MP / MPR chances. Their findings suggest that for shortest itineraries, an extensive criterion in ad-Hoc protocols, bridges between routes lead to longer waiting times and a later re-schedule that increases the delay at the end of the session.

A novel multi-beam directive network MAC protocol utilizing a real-time emulator [43] which concentrates on simultaneous propagation. It will produce a low complexity, distributed, random, MAC access protocol for a multi-beam directional network by utilization of the underlying physical capability of a numerical phased array. In order to ensure the proper power pointing of the transmission beam, the protocol has planned position detection and power management procedures. The presented scheme is also capable of detecting the status of the random-access protocol of a neighbour, meaning that the number of packets lost and interference in the device is significantly decreased. A new Extendable Mobile Network Emulator (EMANE) concept has been put in place to test high reliability in real-time.

In the literature, several models have been developed for the channel assignment problem, e.g. greedy and Tabu-based algorithms [44], the greedy graph-based algorithm [8], and the genetic-algorithm (GA) [45]. Channel assignment model (CAM) is developed for WMN generally classified into two distinct categories: (1) Dynamic model and (2) Quasistatic model [44]. The MRMC (Muti-Radio Multi-Channel) architecture proposed by Liu F, et al. [46] can be used for improvement of the network performance in WMN. The channel assignment method was proposed by Lin J et al. [47] that is dependent on the various parameters viz. window size, etc. In the flow control-oriented channel assignment problem for MRMC, WMNs are collectively given as Mixed Integer Linear Program(MILP) [48]. Avallone S et al. [49] have developed a channel assignment algorithm that depends on the transmission power, allocation of data rate, and traffic flow. Thenral B et al. [50] showed that if the hops count is amplified in WSN, then the network performance diminishes gradually. The summary of the complete literature survey is given in Table 2.

Table 2: Summary of the literature along with pros and cons of the various spectrum and topologies

| Author(s) | Spectrum and Topologies | Protocols | Pros. and Cons. | Remarks |
|---|---|---|---|---|
| [51] | Network topology | Novel neighbor protocol | Polynomial-time (P) algorithm | ROMA protocol greatly increases network efficiency and delay. |
| [11] | Simulation topology | MAC layer protocol, IEEE 802.11 | The creation of spatial channels to increase the throughput | This MAC layer protocol makes use of multiple space division access to receive multiple packages from spatially separated transmitter nodes utilizing the directional reception property. |
| [14] | 100 nodes randomly distributed within a 1500 m* 1500 m network grid for all the simulations | IEEE 802.11 MAC protocol | Mobility is not considered. | The latest smart multi-beam antennas may be widely divided into three types: versatile array antennas, switched multi-beam antennas, and MIMO links |
| [15] | Network topology | Network protocol | Through concurrently distributing many independent data streams over the single path, MIMO links can provide incredibly high spectrally efficient multi-way streaming. | A completely new MAC protocol is needed for the specific features of MIMO links together with many primary optimization considerations. The authors illustrated a centralized algorithm called SCMA, which integrates the core concerns of optimizing the architecture of stream-controlled media access. |
| [16] | Network topology | Routing protocol | With the growing prevalence of local wireless connectivity, there is a strong demand that data sharing among access points and terminals or base stations increase throughput and energy consumption. | Wireless networks are historically configured for single-hop transmission to base stations or to access points via WLAN. But, the increase in wireless network installations has in practice inspired the concept of WMN (WMN) [16], expanding wireless reach, enhancing maximum capacity, and allowing the self-configuration of the network. |
| [17] | Hybrid topology | carrier-sense multiple access (CSMA) protocol | Where multiple accesses are not implemented space-division, the additional beam forming effects can be done easily by optimizing only one antenna unit transmission power. | CSBF requires it to send data that is also used in the supplementary beam. Use this approach to get "broadcast" information to passive nodes within the network by utilizing active nodes to share user-specific data. |
| [18] | Mesh topology | MAC protocol | In order to resolve these difficult issues carefully and improve collaboration efficacy, the authors introduced a new MAC protocol. | Their architecture also takes into account the reverse compatibility, with a multi-beam access point being opened transparently by an IEEE 802.11 terminal. |
| [23] | Network topology | Novel protocol | MAC hybrid (HMAC) protocol's special functionality is not feasible with the conventional MAC protocols for the directional omnidirectional and one ray antennas built on the DCF mechanism, or with the simultaneous transmitting or receiving of the multiple ray antennas. | This article proposes to backward IEEE 802.11 DCF-compatible new MAC hybrid (HMAC) protocol. This protocol allows simultaneous packet receiving (CPR) and simultaneous packet transmission (CPT) on multiple antennas. |
| [27] | Hybrid topology | MAC protocol, IEEE 802.11 | Two reasonable bandwidth assignment issues have been established, which are (a) the Lexicographical Max–Min (LMM) fair Bandwidth Allocation (LMMBA) and (b) the Max-Min equal maximum bandwidth allocation (MMBA) problem. | Linear Programming (LP) for both problems is also recommended for optimal and quick heuristic (MMBA LMMBA) algorithms. Numerical studies have shown that the performance of the LMMBA and MMBA algorithms is very similar to the acceptable maximum performance. |
| [1] | Hybrid topology | MAC protocol, IEEE 802.11 | The implementation of an effective relation planning algorithm is of utmost importance in order to have high throughput for a WMN as observed. | Three channel scheduling algorithms were suggested, based on the spatial TDMA networking network, namely (a) the Network Connection Chart Model, (b) the Network Interference and Noise Ratio Signal, and (c) the Network Chart Model and the receiver threshold conditions. |
| [29] | Mesh topology | MAC protocol, IEEE 802.11 | It is found that interference and colliding substantially reduce mesh network performance using MAC protocols based on containment like IEEE 802.11 DCF. | If transmissions are planned, considerably higher throughput is attainable. The typical methods of estimating optimum schedules are however intractable computationally. |
| [30] | Mesh topology | Routing protocol | The multi-hop WMNs often suffer interference-related connection faults, dynamic barriers, and bandwidth requirement applications. | This interference leads to a significant deterioration in efficiency in WMNs or involves costly manual network management to restore in real-time. |
| [33] | Testbed topology | Routing protocol | A multi-hop MAC Protocol method based on 802.11s for WMNs was submitted. The proposed mesh network is integrated into the connection layer and rather than IP addresses, it is based on its mechanisms on MAC addresses. | This role helps the creation and production of new layer 2 multi-hop CPU-free network systems. The current layer 3 transmission is based on the WMN implementations, since these networks are based on layer 3 transmission based on 802.11 (WLAN), 802.3 (Ethernet), 802.16 (WPAN). |
| [37] | Mesh topology | MAC protocol | The Access Point (AP) and mobile station's finite range is a standard local wireless network. | The AP is typically much more efficient and physically less limited than a Full Function Unit (FFU) mobile station. |

| [38] | Ring and random topology | Threshold access MPC (TAMPC), signal-to-interference ratio (SIR) access MPC (SAMPC) | Explores in IEEE802.11, the asynchronous network with the architectural conditions | The adaptive conditions are adaptive antenna arrays, allowing several peer-to-peer communications simultaneously, and suggest two new access schemes for multi-packet communication (MPC). |

## 4. Performance Metric, Multi-Objective Optimization, and Statistical Testing Framework

Nowadays multi-objective optimization [52] is an important tool for decision-making. It can generate a set of non-dominant (Pareto) solutions which can design a compromise process. The popular evolutionary algorithms viz. genetic algorithm (GA) [52–54] enable solving the single-objective and multi-objective optimization problems in the polynomial time complexity to attain near-global optimal results. Likewise, the inheritance-based GA is also described in the literature [55]. These are easy to apply to continuous search space problems.

The different models to address different issues and challenges in the MAC layer are developed by researchers in this area. These models are compared by using different quality of service (QoS) parameters e.g. throughput, bandwidth, etc. A few QoS parameters are as follows.

$$Bandwidth\ (BW) = (f_{max} - f_{min}) \qquad (1)$$

where, $f_{max}$ and $f_{min}$ are maximum and minimum value of the frequencies of the system respectively.

Bandwidth is considered as a potential link measurement. It is a very important QoS parameter which may be applied to facilitate different user demands in different format.

Throughput is the actual volume of data sent/received without any failure over the communication channel. In other words, it is a true measurement of how fast one can send data across a network.

The total delay/latency of a network can be calculated as follows.

$$\begin{aligned} Total\ Delay\ (Dt) &= Transmission\ Delay\ (Dtrans) + Propagation\ Delay\ (Dprop) \\ &+ Queuing\ Delay\ (Dq) \\ &+ Processing\ Delay(Dproc) \end{aligned} \qquad (2)$$

$$Transmission\ Delay\ (Dtrans) = \frac{Message\ size}{Bandwidth} \qquad (3)$$

$$Propagation\ Delay\ (Dprop) = \frac{Distance}{Transmission\ Speed} \qquad (4)$$

$$Queuing\ Delay\ (Dq) = \frac{(no.\ of\ data\ packets\ - 1)\ Message\ size}{2\ Bandwidth} \qquad (5)$$

These objectives should be minimized during the design and analysis of the multi-objective optimization.

The **availability** (*A*) can be computed by a proportion of the expected value of the uptime of a system (EUTS) to the total value of the expected up and down time.

$$A = \frac{expected\ value\ of\ the\ uptime\ of\ a\ system\ (EUTS)}{total\ amount\ of\ time\ (T)} \qquad (6)$$

Another equation to compute **availability** (*A*) is a ratio of the Mean Time to Failure (MTTF) and sum of the MTTF and Mean Time to Repair (MTTR).

$$A = \frac{MTTF}{MTTF + MTTR} \qquad (7)$$

Packet Delivery Ratio (**PDR**) is calculated using the below formula, where *DR* is Delivery Ration and *DL* is Delivery Length.

$$PDR = \frac{DR}{DR + DL} \tag{8}$$

These objectives should be maximized during the design and analysis of multi-objective optimization.

The occurrence analysis of QoS parameters in the different techniques and protocols in the literature under consideration is shown in Tables 3-5. Table 3 lists such techniques and protocols along with number of QoS parameters considered in each one. Table 4 reveals the name of QoS parameters used in protocols and the value of these parameters for different protocols are given in Table 5.

Table 3: Number of QoS Parameters in the different Protocols in Literature under the study

| Techniques/Protocols | Number of QoS Parameters |
|---|---|
| Novel neighbour protocol | 2 |
| Network protocol | 1 |
| MAC layer protocol (IEEE 802.11) | 3 |
| Routing protocol | 4 |
| CSMA protocol | 3 |
| MAC protocol | 3 |
| Novel protocol (Hybrid MAC) | 2 |
| Novel MAC protocol (TAMPC, SAMPC) | 2 |
| Directional MAC | 1 |
| MACA/MACAW protocol | 2 |
| Novel transport layer protocol | 3 |
| Ad-Hoc transport protocol (ATP) | 2 |
| Live scheduling algorithm | 1 |
| MB-URAM | 2 |
| CBPFS | 2 |
| MMBA & LMMBA algorithm | 3 |
| CBF (complementary beamforming) | 2 |
| Minimum interference channel assignment algorithm (MICA) | 2 |
| Heuristic channel assignment algorithm | 1 |
| Mixed integer linear program (MILP) | 2 |
| Angle based multicast routing algorithm (AMRA) | 2 |

Table 4: QoS Parameters used in Protocols

| Techniques/Protocols | QoS Parameters | | | | | |
|---|---|---|---|---|---|---|
| | Bandwidth | Delay | PDR | Throughput | Fairness | Energy Consumption |
| Novel neighbour protocol | | ✓ | | ✓ | | |
| Network protocol | | | | ✓ | | |
| MAC layer protocol (IEEE 802.11) | | | ✓ | ✓ | ✓ | |
| Routing protocol | ✓ | ✓ | | ✓ | ✓ | |
| CSMA protocol | | ✓ | ✓ | ✓ | | |
| MAC protocol | | ✓ | | ✓ | ✓ | |
| Novel protocol (Hybrid MAC) | | ✓ | | ✓ | | |
| Novel MAC protocol (TAMPC, SAMPC) | | | | ✓ | ✓ | |
| Directional MAC | | | | ✓ | | |
| MACA/MACAW protocol | | | ✓ | ✓ | | |
| Novel transport layer protocol | ✓ | ✓ | ✓ | | | |
| Ad-hoc transport protocol (ATP) | | | | ✓ | ✓ | |
| Live scheduling algorithm | | | | ✓ | | |
| MB-URAM | ✓ | | | | | ✓ |
| CBPFS | | | | ✓ | ✓ | |
| MMBA & LMMBA algorithm | ✓ | | | ✓ | ✓ | |
| CBF (complementary beam-forming) | | ✓ | | ✓ | | |
| Minimum interference channel assignment algorithm (MICA) | | ✓ | | ✓ | | |
| Heuristic channel assignment algorithm | | | | ✓ | | |
| Mixed integer linear program (MILP) | | ✓ | | ✓ | | |
| Angle based multicast routing algorithm (AMRA) | | | ✓ | ✓ | | |

Table 5: Value of QoS Parameters used in Protocols

| Techniques/Protocols | QoS Parameters | | | | | |
|---|---|---|---|---|---|---|
| | Bandwidth | Delay | PDR | Throughput | Fairness | Energy Consumption |
| Novel neighbour protocol | | 1 | | 1 | | |
| Network protocol | | | | 1 | | |
| MAC layer protocol (IEEE 802.11) | | | 1 | 4 | 1 | |
| Routing protocol | 1 | 2 | | 2 | 1 | |
| CSMA protocol | | 2 | 1 | 2 | | |
| MAC protocol | | 2 | | 6 | 1 | |
| Novel protocol (Hybrid MAC) | | 1 | | 1 | | |
| Novel MAC protocol (TAMPC, SAMPC) | | | | 1 | 1 | |
| Directional MAC | | | | 1 | | |
| MACA/MACAW protocol | | | 1 | 1 | | |
| Novel transport layer protocol | 1 | 1 | 1 | | | |
| Ad-hoc transport protocol (ATP) | | | | 1 | 1 | |
| Live scheduling algorithm | | | | 1 | | |
| MB-URAM | 1 | | | | | 1 |
| CBPFS | | | | 1 | 1 | |
| MMBA & LMMBA algorithm | 1 | | | 1 | 1 | |
| CBF (complementary beam-forming) | | 1 | | 1 | | |
| Minimum interference channel assignment algorithm (MICA) | | 1 | | 1 | | |
| Heuristic channel assignment algorithm | | | | 1 | | |
| Mixed integer linear program (MILP) | | 1 | | 1 | | |
| Angle based multicast routing algorithm (AMRA) | | | 1 | 1 | | |

The significance-based testing is the utmost common framework designed for statistical hypothesis testing (SHT) [56, 57]. However, alternative testing is needed to assess a set of statistical methods.

Purpose: To detect the significance between two population means, paired- difference design.

$H_0$: The population average is same.   $H_a$: The population average is not same.

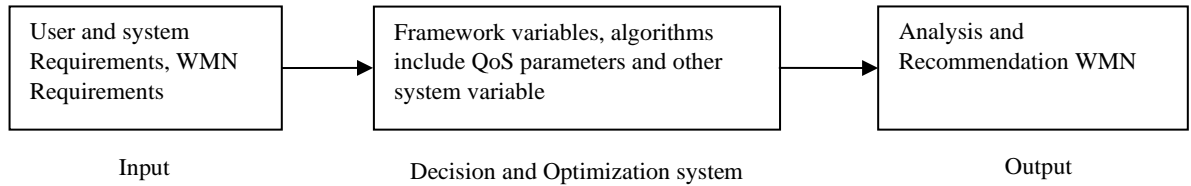

Input    Decision and Optimization system    Output

Figure 3: Block diagram of the proposed future framework for critical analysis and recommendation

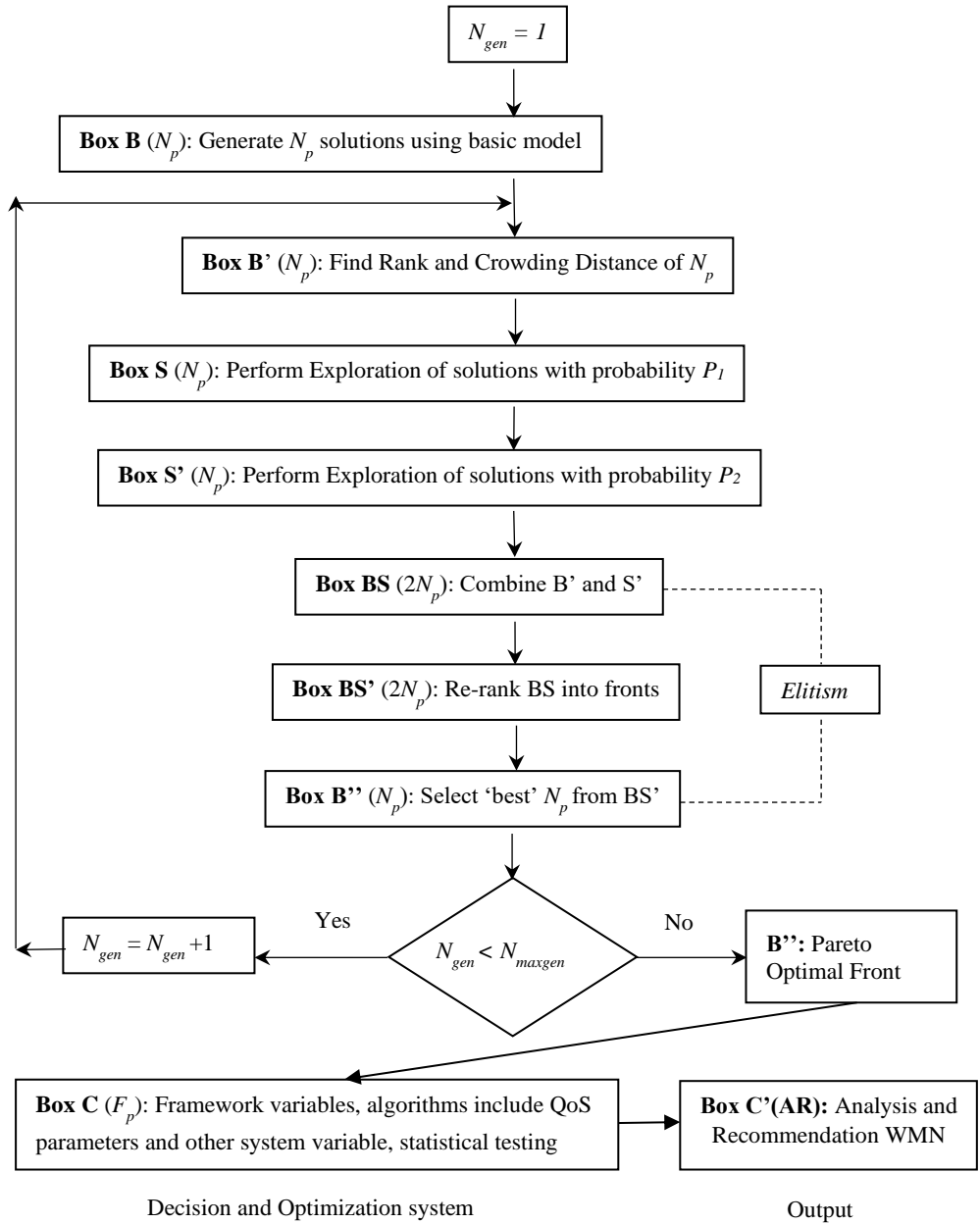

Figure 4: Flowchart of Suggested Multi-Objective Statistical framework

The framework given in Figures 3-4 will take as the QoS parameters for MAC Protocols then apply the MO technique and then apply STH. In this way, it will generate the recommendation for the WMN.

## 5. Conclusion and Future Work

A detailed literature review of the WMN and IEEE 802.11 research issues are discussed in this paper. After the literature survey in various aspects and fields, it is observed that the main factors that should be focused on for MAC protocol for an efficient WMN are topology, architecture, data pattern, traffic, number of channels applied by different nodes, density of the nodes, mobility of the nodes, transmission power, routing, scalability, security, and interoperability which improve the QoS in WMN. Further, critical comparative analysis and recommendation in MAC protocols for WMN using multi-objective optimization and statistical testing framework are performed. This framework is used for the analysis and recommendation of different protocols available for the QoS Parameters. In future work, we will study the proposed framework as a use case, and also emphasize the enhancements of this framework. Furthermore, future work is also possible for security-guided algorithms with different constraints for this purpose to satisfy the user demand in the current scenario.

**Declarations**

**Authors' contributions:** Conceptualization: Ankita Sing, Sudhakar Sing; Shiv Prakash; Methodology: Ankita Sing, Sudhakar Sing, Shiv Prakash; Formal analysis and investigation: Ankita Singh, Sudhakar Singh, Shiv Prakash; Writing - original draft preparation: Ankita Singh, Sudhakar Singh; Writing - review and editing: Sudhakar Singh, Shiv Prakash; Resources: Ankita Singh, Sudhakar Singh, Shiv Prakash; Supervision: Sudhakar Singh, Shiv Prakash.

**Conflicts of interest:** The authors declare that they have no conflict of interest in this paper.

**Funding:** No funding was received for conducting this study.

**Availability of data and material:** Not applicable.

**Code availability:** Not applicable.

**References**

1. Gore AD, Karandikar A (2011) Link scheduling algorithms for wireless mesh networks

2. Zhang H, Wu S, Zhang C, Krishnamoorthy S (2021) Optimal Distribution in Wireless Mesh Network with Enhanced Connectivity and Coverage. In: Advances in Intelligent Systems and Computing. pp 1117–1128

3. Goldberg BS, Hall JE, Pham PK, Cho CS (2021) Text messages by wireless mesh network vs voice by two-way radio in disaster simulations: A crossover randomized-controlled trial. Am J Emerg Med

4. Rao AN, Babu PR, Reddy AR (2021) Analysis of Wireless Mesh Networks in Machine Learning Approaches. In: Proceedings of International Conference on Advances in Computer Engineering and Communication Systems. pp 321–331

5. Rajendran R, others (2021) An optimal strategy to countermeasure the impersonation attack in wireless mesh network. Int J Inf Technol 1–6

6. Nsaif SA, Park SY, Rhee JM (2021) SRAD: A Novel Approach to Seamless Routing for Wireless Ad Hoc Networks. In: 2021 23rd International Conference on Advanced Communication Technology (ICACT). pp 172–175

7. Hosseinabadi G, Vaidya N (2013) Token-DCF: An opportunistic MAC protocol for wireless networks

8. Marina MK, Das SR, Subramanian AP (2010) A topology control approach for utilizing multiple channels in multi-radio wireless mesh networks. Comput Networks 54:241–256. https://doi.org/10.1016/j.comnet.2009.05.015

9. Bharghavan V, Demers A, Shenker S, Zhang L (1994) Macaw. ACM SIGCOMM Comput Commun Rev 24:212–225. https://doi.org/10.1145/190809.190334

10. Cooper M, Goldburg M (1996) Intelligent antennas: Spatial division multiple access. Annu Rev Commun 4:2–13

11. Lal D, Toshniwal R, Radhakrishnan R, et al (2002) A novel MAC layer protocol for space division multiple access in wireless ad hoc networks

12. Ramanathan R (2005) Antenna beamforming and power control for ad hoc networks. Mob Ad Hoc Netw 139–174. https://doi.org/10.1002/0471656895.ch5

13. Lal D, Jain V, Zeng QA, Agrawal DP (2004) Performance evaluation of medium access control for multiple-beam antenna nodes in a wireless LAN. IEEE Trans Parallel Distrib Syst 15:1117–1129. https://doi.org/10.1109/TPDS.2004.84

14. Sundaresan K, Anantharaman V, Hsieh H-Y, Sivakumar R (2003) ATP. In: Proceedings of the 4th ACM international symposium on Mobile ad hoc networking & computing - MobiHoc '03. ACM Press, New York, New York, USA, p 64

15. Sundaresan K, Sivakumar R (2004) A unified MAC layer framework for ad-hoc networks with smart antennas. In: Proceedings of the International Symposium on Mobile Ad Hoc Networking and Computing (MobiHoc). Association for Computing Machinery, pp 244–255

16. Akyildiz IF, Wang X, Wang W (2005) Wireless mesh networks: A survey. Comput Networks 47:445–487. https://doi.org/10.1016/j.comnet.2004.12.001

17. Yang-Seok Choi, Alamouti SM, Tarokh V (2006) Complementary beamforming: new approaches. IEEE Trans Commun 54:41–50. https://doi.org/10.1109/TCOMM.2005.861674

18. Wang G, Xiao P, Li W (2017) A novel MAC protocol for wireless network using multi-beam directional antennas. 2017 Int Conf Comput Netw Commun ICNC 2017 36–40. https://doi.org/10.1109/ICCNC.2017.7876098

19. Choudhury RR, Yang X, Ramanathan R, Vaidya NH (2002) Using directional antennas for medium access control in ad hoc networks. In: Proceedings of the Annual International Conference on Mobile Computing and Networking, MOBICOM. Association for Computing Machinery (ACM), pp 59–70

20. Gossain H, Cordeiro C, Cavalcanti D, Agrawal DP (2004) The deafness problems and solutions in wireless ad hoc networks using directional antennas. GLOBECOM - IEEE Glob Telecommun Conf 108–113. https://doi.org/10.1109/glocomw.2004.1417558

21. Takata M, Bandai M, Watanabe T (2007) A MAC protocol with directional antennas for deafness avoidance in ad hoc networks

22. Choudhury RR, Vaidya NH (2004) Deafness: A MAC problem in ad hoc networks when using directional antennas

23. Jain V, Gupta A, Agrawal DP (2008) On-demand medium access in multihop wireless networks with multiple beam smart antennas. IEEE Trans Parallel Distrib Syst 19:489–502. https://doi.org/10.1109/TPDS.2007.70739

24. Shihab E, Cai L, Pan J (2009) A distributed asynchronous directional-to-directional MAC protocol for wireless ad hoc networks. IEEE Trans Veh Technol 58:5124–5134. https://doi.org/10.1109/TVT.2009.2024085

25. Akyildiz IF, Lee WY, Chowdhury KR (2009) CRAHNs: Cognitive radio ad hoc networks. Ad Hoc Networks 7:810–836. https://doi.org/10.1016/j.adhoc.2009.01.001

26. Conti M, Giordano S (2007) Multihop ad hoc networking: The reality. IEEE Commun Mag 45:88–95. https://doi.org/10.1109/MCOM.2007.343617

27. Tang J, Hincapié R, Xue G, et al (2010) Fair bandwidth allocation in wireless mesh networks with cognitive radios. IEEE Trans Veh Technol 59:1487–1496. https://doi.org/10.1109/TVT.2009.2038478

28. Liu T, Liao W (2010) Multicast routing in multi-radio multi-channel wireless mesh networks. IEEE Trans Wirel Commun 9:3031–3039. https://doi.org/10.1109/TWC.2010.082310.090568

29. Wang P, Bohacek S (2011) Practical computation of optimal schedules in multihop wireless networks

30. Kim KH, Shin KG (2011) Self-reconfigurable wireless mesh networks. IEEE/ACM Trans Netw 19:393–404. https://doi.org/10.1109/TNET.2010.2096431

31. Liu E, Zhang Q, Leung KK (2011) Clique-based utility maximization in wireless mesh networks. IEEE Trans Wirel Commun 10:948–957. https://doi.org/10.1109/TWC.2011.011111.100790

32. Hiertz G, Mingozzi E, Serrano P (2011) Workshop MeshTech 2011 Enabling Technologies and Standards for Wireless Mesh Networking. 2011

33. Carrano RC, Magalhães LCS, Saade DCM, Albuquerque CVN (2011) IEEE 802.11s multihop MAC: A tutorial. IEEE Commun Surv Tutorials 13:52–67. https://doi.org/10.1109/SURV.2011.040210.00037

34. Huang Y, Yang X, Yang S, et al (2011) A cross-layer approach handling link asymmetry for wireless mesh access networks. IEEE Trans Veh Technol 60:1045–1058. https://doi.org/10.1109/TVT.2011.2106172

35. Aziz A, Starobinski D, Thiran P (2011) Understanding and tackling the root causes of instability in wireless mesh networks

36. Bazan O, Jaseemuddin M (2012) A survey on MAC protocols for wireless adhoc networks with beamforming antennas. IEEE Commun Surv Tutorials 14:216–239. https://doi.org/10.1109/SURV.2011.041311.00099

37. Chou ZT, Huang CQ, Chang JM (2014) QoS provisioning for wireless LANs with multi-beam access point. IEEE Trans Mob Comput 13:2113–2127. https://doi.org/10.1109/TMC.2013.85

38. Babich F, Comisso M, Crismani A, Dorni A (2015) On the design of MAC protocols for multi-packet communication in IEEE 802.11 heterogeneous networks using adaptive antenna arrays. IEEE Trans Mob Comput 14:2332–2348. https://doi.org/10.1109/TMC.2014.2385058

39. Kuperman G, Margolies R, Jones NM, et al (2016) Uncoordinated MAC for Adaptive Multi-Beam Directional Networks: Analysis and Evaluation

40. Wang Y, Wu Y, Wang Y, et al (2017) Antioxidant properties of probiotic bacteria. Nutrients 9:. https://doi.org/10.3390/nu9050521

41. Hong W, Jiang ZH, Yu C, et al (2017) Multibeam Antenna Technologies for 5G Wireless Communications. IEEE Trans Antennas Propag 65:6231–6249. https://doi.org/10.1109/TAP.2017.2712819

42. Medjo Me Biomo JD, Kunz T, St-Hilaire M (2018) Exploiting Multi-Beam Antennas for End-to-End Delay Reduction in Ad Hoc Networks. Mob Networks Appl 23:1293–1305. https://doi.org/10.1007/s11036-018-1037-8

43. Proulx B, Madiedo J, Jones NM, Kuperman G (2018) Topology control in aerial multi-beam directional networks

44. Subramanian AP, Gupta H, Das SR, Cao J (2008) Minimum interference channel assignment in multiradio wireless mesh networks. IEEE Trans Mob Comput 7:1459–1473. https://doi.org/10.1109/TMC.2008.70

45. Jia J, Chen J, Chang G, Tan Z (2009) Energy efficient coverage control in wireless sensor networks based on multi-objective genetic algorithm. Comput Math with Appl 57:1756–1766. https://doi.org/10.1016/j.camwa.2008.10.036

46. Liu F, Bai Y (2012) An overview of topology control mechanisms in multi-radio multi-channel wireless mesh networks. EURASIP J Wirel Commun Netw 2012:324. https://doi.org/10.1186/1687-1499-2012-324

47. Lin JW, Lin SM (2014) A weight-aware channel assignment algorithm for mobile multicast in wireless mesh networks. J Syst Softw 94:98–107. https://doi.org/10.1016/j.jss.2014.03.040

48. Franklin AA, Bukkapatanam V, Murthy CSR (2011) On the end-to-end flow allocation and channel assignment in multi-channel multi-radio wireless mesh networks with partially overlapped channels q. Comput Commun 34:1858–1869. https://doi.org/10.1016/j.comcom.2011.05.006

49. Avallone S, Stasi G Di (2013) An experimental study of the channel switching cost in multi-radio wireless mesh networks. IEEE Commun Mag 51:124–134. https://doi.org/10.1109/MCOM.2013.6588661

50. Thenral B, Thirunadana Sikamani K (2015) AMRA: Angle based multicast routing algorithm for wireless mesh networks. Indian J Sci Technol 8:974–6846. https://doi.org/10.17485/ijst/2015/v8i13/59451

51. Bao L, Garcia-Luna-Aceves JJ (2002) Transmission scheduling in ad hoc networks with directional antennas. In: Proceedings of the Annual International Conference on Mobile Computing and Networking, MOBICOM. Association for Computing Machinery (ACM), pp 48–58

52. Deb K (2001) Multi-Objective Optimization Using Evolutionary Algorithms. John Wiley Sons, New York, USA

53. Goldberg DE, Holland JH (1988) Genetic Algorithms and Machine Learning. Mach Learn 3:95–99. https://doi.org/10.1023/A:1022602019183

54. Rangaiah GP (2009) Multi-Objective Optimization: Techniques and Applications in Chemical Engineering (Advances in Process Systems Engineering)

55. Kumar M, Guria C (2017) The elitist non-dominated sorting genetic algorithm with inheritance (i-NSGA-II) and its jumping gene adaptations for multi-objective optimization. Inf Sci (Ny). https://doi.org/10.1016/j.ins.2016.12.003

56. Sohn W, Hong E (2021) Monte Carlo simulation for verification of nonparametric tests used in final status surveys of MARSSIM at decommissioning of nuclear facilities. Nucl Eng Technol 53:1664–1675

57. Trivedi V, Prakash S, Ramteke M (2017) Optimized on-line control of MMA polymerization using fast multi-objective DE. Mater Manuf Process 32:1144–1151